\documentclass[preprint,preprintnumbers,amsmath,amssymb]{revtex4}
\usepackage{graphicx}
\usepackage{graphics}
\usepackage{amssymb}

\newcommand{\bm}[1]{\mbox{\boldmath $#1$}}     
\newcommand{\mfrac}[2]{{\textstyle\frac{#1}{#2}}}

\newcommand{\uniC}{\mbox{i}}
\def\re{\mathbb{R}}
\newcommand{\dpar}[2]{\frac{\partial #1}{\partial #2}}

\newcommand{\diag}{\mbox{diag\,}}

\begin{document}

\title{Numerical treatment of the light propagation problem in the post-Newtonian formalism}

\author{A. San Miguel, F. Vicente and J.-F. Pascual-S\'anchez}

\address{Dept. de Matem\'atica Aplicada,
Facultad de Ciencias.\\
Universidad de Valladolid, 47005 Valladolid, Spain}


\begin{abstract}
The geometry of a light wavefront, evolving from a initial flat wavefront in the 3-space associated with a post-Newtonian relativistic spacetime,  is studied numerically by means of the ray tracing method. For a discretization of the bidimensional light wavefront, a surface fitting technique is used to determine the curvature of this surface. The relationship between the intrinsic curvature of the wavefront and the change of the arrival time at different points on the Earth is also numerically discussed.
\end{abstract}
\maketitle

\section{Introduction}
\vspace{.2cm}
According to Einstein's theory of general relativity, a generic gravitational field has, at a time, three main effects on light:  the Shapiro time delay, the bending
of rays and the curvature of wavefronts. The gravitational field acts respectively as a retarder delaying  a light wavefront, as a prism tilting  a  light wavefront, and as a lens curving a light wavefront. In special relativity, time delays  can be also induced by inertial motion of the observer or the source as in the Doppler effect and uniform motion of a observer can result in aberration, i.e., an apparent change in the position of the source.

Hence, the curvature of a initially plane light wavefront by a gravity field  is a purely general relativistic effect that has no special relativistic analogue.
In order to obtain an experimental measurement of the curvature of a wave front, Samuel~\cite{Sam}
proposed a method  based on the relation between the differences of arrival time recorded at four points on the Earth and the volume of a parallelepiped determined by four points in the curved wavefront surface. We will establish a discretized model of the wavefront surface by means of a regular triangulation for the study of the curvature(s) of this surface.
The main  methods and results have been recently  published in \cite{san}, in this work we will only expose a summary of them.

\section{Light propagation in a gravitational field}
\vspace{.2cm}
Let us consider a spacetime $(\mathcal{M},g)$ corresponding to a weak gravitational field determined by a metric tensor given in a global coordinate system  $\{(\bm{z},ct)\}$ by
    \medskip
        \begin{equation}
                g_{\alpha\beta}=\eta_{\alpha\beta}+h_{\alpha\beta}, \qquad {\rm{with}}\quad \eta_{\alpha\beta}=\diag(1,1,1,-1).
        \end{equation}
            where  the coordinate components of the metric deviation $h_{\alpha\beta}$ are: \medskip
        \begin{equation}
            h_{ab} =  2c^{-2}\kappa \|\bm{z}\|^{-1}\delta_{ab},\quad
            h_{a4} = -4c^{-3}\kappa \|\bm{z}\|^{-1}\dot{Z}_a, \quad
            h_{44} =  2c^{-2}\kappa \|\bm{z}\|^{-1},
        \end{equation}

        here $\kappa:=GM$ represents the gravitational constant of the Sun, located at $Z^a(t)$,
         and \,$c$\, represents the vacuum light speed.

The null geodesics, $z(t)=\big(\bm{z}(t),t\big)$ satisfy the following equations (see \cite{Bru}):
    \begin{eqnarray*}
          \ddot{z}^a & =& \mfrac{1}{2} c^2h_{44,a} -[\mfrac{1}{2}h_{44,t}\delta^a_k+h_{ak,t}+c(h_{4a,k}-h_{4k,a})]\dot{z}^k\nonumber\\
        & & -(h_{44,k}\delta^a_l+h_{ak,l}-\mfrac{1}{2}h_{kl,a})\dot{z}^k\dot{z}^l \nonumber\\
        & & -(c^{-1}h_{4k,j}-\mfrac{1}{2}c^{-2}h_{jk,t})\dot{z}^j\dot{z}^k\dot{z}^a,\label{PN1a}\\[.1in]
0 & = &  g_{\alpha\beta}\dot{z}^\alpha\dot{z}^\beta.\label{PN1b}
\end{eqnarray*}
   where, in the first equation, the first and third terms in the right are of order $O(1)$, while the remaining terms are $O(c^{-1})$. The second equation is the isotropy constraint satisfied by the null geodesics.

\section{Numerical description of a spacelike bidimensional light wavefront}

\vspace{.2cm}
\subsection{Discretization of the initial wavefront}

We consider a flat initial surface $\mathcal{S}_0$ far from the Sun formed by points $(z_1,z_2,-\zeta)$ (with $\zeta>0$) in an asymptotically Cartesian coordinate system $\{z\}$. For the discretization of $\mathcal{S}_0$ a triangulation is constructed in such a form that each vertex is represented by a complex number of the  set:
        \begin{equation}\label{vertices1}
            \bar{\mathcal{V}}:=\{z=a_1+a_2\omega+a_3\omega^2\;|\;\; a_1,a_2,a_3\in\mathcal{A},\; \omega:=\exp(2\pi \uniC/3)\},
        \end{equation}
     The complex plane and the plane $\mathcal{S}_0$ may be identified by means of the mapping $\iota:z\mapsto (\Re(z),\Im(z),-\zeta)$. Thus a regular triangulation $\mathcal{V}$ of the initial wavefront $\mathcal{S}_0$ is determined. At each vertex in $\mathcal{V}$ a photon $\bm{z}_0:=\bm{z}(0)\in\mathcal{V}$ with velocity $\dot{\bm{z}}_0:=(0,0,c)$ is located. The null geodesics equation may be written as a {first order} differential system {$\dot{\bm{u}}=\bm{F}(\bm{u},t)$}, in phase space $\bm{u}=(\bm{z},\dot{\bm{z}})$, which determines
in $\mathcal{E}$ (the quotient space of $ \mathcal{M} $ by the global timelike vector field  $ \partial_t$ associated to the global coordinate system used in the post-Newtonian formalism) {a flow}: $\bm{z}(t) =\varphi_t(\bm{z}_0,\dot{\bm{z}}_0)$.
For each time $t$ the image of $\mathcal{S}_0$ under the flow $\Phi_t$ determines a curved wavefront $\mathcal{S}_t$.
The initial triangulation by $\mathcal{V}$ induces a triangulation on the final wavefront $\mathcal{S}_t$,  whose vertices we enumerate using the same labels used for the corresponding vertices in $\mathcal{S}_0$.\\

Let  $\{y^j\}_{j=1}^3$ be a \emph{normal coordinate system} with pole at the point $P\in \mathcal{S}_t$ and associated normal reference frame $\{\bm{e}_i\}_{i=1}^3$. Under the coordinate transformation $z^i\mapsto y^i $, from post-Newtonian to normal coordinates, the metric tensor $\gamma$ on the space $\mathcal{E}$  is determined (up  to terms of first order in $\tilde{\Gamma}$) from $\tilde{\gamma}_{ab}:=g_{ab}-\mfrac{g_{a4}g_{b4}}{g_{44}}$ by
\begin{equation}\label{trimetricNor}
    \gamma_{ij} =\dpar{z^a}{y^i}\dpar{z^b}{y^j}\tilde{\gamma}_{ab} \quad
{\rm where} \quad
    z^a= a_0^a+\Lambda^a_i\big(y^i - \mfrac{1}{2}\big(\tilde{\Gamma}^i_{jk}\big)_0y^jy^k\big).
\end{equation}

\subsection{Local approximation of the wavefront}%

To compute differential magnitudes of the wavefront surface corresponding to the mesh $\mathcal{V}$
at each inner vertex $z_j^{0}\in\bar{\mathcal{V}}, j=1,\dots,J$ we consider the 1--ring $[z^0_j;z^0_{j_k}]_{j=1,\dots,J,k=0,\dots,5}$ formed by the six vertices $z^0_{j_k}$ closest to $z^0_j$:
For each 1--ring $[z^*_j;z^*_{j_k}]$ one obtains on the mesh $\mathcal{V}$ the image 1--ring $[z_j;z_{j_k}]$ under the flow $\varphi_t$.

In a neighbourhood of the image point $z_j$ the wavefront can be approximated by a least-squares fitting of the data $[z_j;z_{j_k}]$ as the quadric $\bar{\mathcal{S}}_t$:
        \begin{equation}\label{2}
            y^3=f(y^1,y^2):=\mfrac{1}{2}a_1(y^1)^2 + a_2y^1 y^2 + \mfrac{1}{2}a_3(y^2)^2.
        \end{equation}
        using adapted normal coordinates $\{y^i\}$.

\subsection{Curvature of the wavefront}

Using coordinates $x^1,x^2$ adapted to the quadric $\bar{\mathcal{S}}_t$:
the metric $\bm{\gamma}$ on $\mathcal{E}$ induces a metric on $\tilde{\mathcal{S}}_j$ of the form
        \begin{equation}\label{metrind}
            g_{AB} := \gamma_{ij}\dpar{y^i}{x^A}\dpar{y^j}{x^B}.
        \end{equation}
     By means of a generalized Gauss formula, the difference of sectional curvatures $K$ and $\bar{K}$ associated with the plane generated by the tangent vectors $\{\bm{v}_1,\bm{v}_2\}$, in $\mathcal{S}$ and $\mathcal{E}$ respectively, is the  { relative sectional curvature $K_{\rm{rel}}$} (see ~\cite{dC}):
        \begin{equation}\label{FGt}
       {    K_{\rm{rel}}:=K(\bm{v}_1,\bm{v}_2)-\bar{K}(\bm{v}_1,\bm{v}_2)=\lambda_1\lambda_2,}
        \end{equation}
        whereas the  { mean curvature $H$} is determined by half the trace of the second fundamental form $I\!I$:
        \begin{equation}\label{FGm}
{  H=\mfrac{1}{2}(\lambda_1+\lambda_2)}
        \end{equation}

\section{Application of a numerical integrator to the case of a static Sun}
\vspace{.2cm}
The ray tracing has been carried out using an integrator based on the classic Taylor series method for ordinary differential equations (see \cite{JZ}).
This integrator presents the following advantages: allows the control of both the order and the step size employed in the method. In this integrator one may use extended arithmetic precision  for the highly accurate computation required in this problem. (Precision of $120$ binary digits and a tolerance $\rm{\tt Tol}=1.{\rm E} -20$ are used to solve this problem.) At each step the null constraint equation is preserved.\\

Hereafter, we consider the simplest gravitational model generated by a static Sun, considered as a point. However, it can be applied to more complex and realistic gravitational fields of the solar system, as considered in   \cite{KP}.
We now apply the ray tracing method   to a light wavefront region propagating along a tubular neighborhood around the $Oz^3$-axis and where the inner and outer hexagons have radii of lengths {$R_\odot/25$} and  {$2R_\odot$}, respectively. The initial flat wave surface {$\mathcal{S}_0$}, perpendicular to the $Oz^3$-axis, is positioned at {-100 A.U.}. The wavefront {$\mathcal{S}_T$} is then determined after a trip of {101 A.U.}, i.e., at the position of the Earth. Remember that the focal length for a light ray grazing an opaque Sun is located at 548.30 A.U. from the Sun.\\

\begin{figure}[h]
\begin{center}
\includegraphics[width=8cm,angle=0]{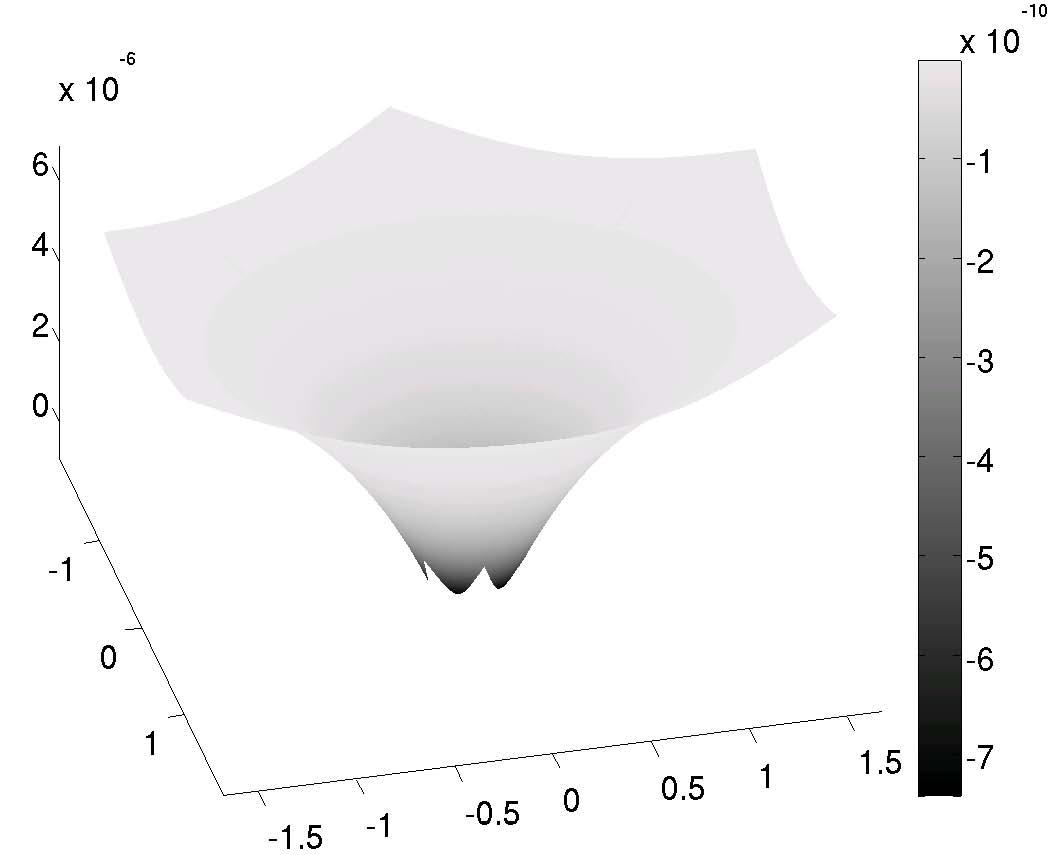}
\caption{Wavefront surface and relative sectional curvature (gray scale) deformed by a spherical gravitational field (a different scale is used for the vertical axis).}\label{figure}
\end{center}
\end{figure}

In Figure \ref{figure}, the surface $\mathcal{S}_T$ at the time when the wavefront arrives at the Earth is shown using a gray-scale to represent the relative sectional curvature (note we have used a different scale on the $Oz^3$--axis). One sees in this figure that the absolute value of the relative sectional curvature function defined on $\mathcal{S}_T$ increases as the distance between the photon and the $Oz^3$--axis decreases.

\section{Variation of the time of arrival and curvature of the light wavefront}
\vspace{.2cm}
Suppose four receiving stations located at four points on an Earth  hemisphere. The arrival time differences between these stations depend on the curvature of the wavefront.  Assumed known the measurements of arrival times corresponding to four points $\{Q_a\}_{a=1}^4$ on the wavefront it is possible to determine an approximation of the wavefront curvature in a region far enough from the Sun (say the Earth), without resorting to the ray tracing method.

An estimation of the Gaussian curvature of the wavefront surface can be obtained using the notion of the Wald curvature  of a metric space established in Distance Geometry (see \cite{Blu}) that in the case of 2-dimensional manifolds  agrees with the Gaussian curvature. The Wald curvature is determined  as the limit of the embedding curvatures of metric quadruples isometrically embedded in  surfaces of constant curvature (the Euclidean plane $\re^2$, the 2--sphere $\mathbb{S}^2_{\sqrt{\kappa}}$  or  the hyperbolic space $\mathbb{H}^2_{\sqrt{-\kappa}}$).\\

In the hyperbolic plane $\mathbb{H}^2_r$ of curvature $-1/r^2$, represented by the Blumenthal model (\cite{Blu}) we consider the metric quadruple  associated  with the points $\{Q_a\}$ assuming that the geometry of the 3--space in the vicinity of the Earth is Euclidean.

The Wald curvature associated to the chosen  quadruple $\{Q_a\}$  prove to be
\begin{equation}
\kappa := -\mfrac{1}{r^2}= {-0.16}\, {\rm E}{-10}.
\end{equation}
This result gives an approximation of the total curvature of the wavefront surface under the assumption that locally this surface may be identified with a hyperbolic plane in which the quadruple considered is isometrically embedded.

\acknowledgements

This research was partially supported by the Spanish Ministerio de Educaci\'on y Ciencia, MEC-FEDER grant ESP2006-01263.



\begin{thebibliography}{9}
\bibitem{Sam} Samuel J 2004 {\it Class. Quantum Grav.\/}, {\bf 21}, L83--L88.
\bibitem{san} San Miguel A, Vicente F and Pascual-S\'anchez J F 2009 {\it Class. Quantum Grav.\/}, {\bf 26}, 235004.
\bibitem{Bru} Brumberg V A 1991 {\it Essential Relativistic Celestial Mechanics}, (Bristol: Adam Hilger).
\bibitem{dC} do Carmo M P 1992 {\it Riemannian Geometry} (Boston: Birkh\"auser).
\bibitem{JZ} Jorba \`A and Zou M 2005 {\it Experimental Mathematics} {\bf 14}, 99-117.
\bibitem{KP} Klioner S A and Peip M 2003 {\it Astron. Astrophys.\/} {\bf 410}, 1063--1074.
\bibitem{Blu} Blumenthal L M 1970 {\it Theory and Applications of Distance Geometry}. (New York: Chelsea Publishing Company, 2nd edition).
    \end{thebibliography}
\end{document}